\documentclass[12pt]{article}
\usepackage{graphicx}
\usepackage{amsmath,amssymb}
\newcommand{\bc}{\begin{center}}
\newcommand{\ec}{\end{center}}
\newcommand{\be}{\begin{equation}}
\newcommand{\ee}{\end{equation}}
\newcommand{\bea}{\begin{eqnarray}}
\newcommand{\eea}{\end{eqnarray}}
\newcommand{\bsl}{\boldsymbol}
\begin{document}
\begin{center}
{\Large\bf \boldmath Heavy baryon spectroscopy in the QCD string
model }\footnote{~~Contribution at PANIC08, Eilat, Israel,
November
2008.} 

\vspace*{6mm}
{I.~L.~Grach, I.~M.~Narodetskii, M.~A.~Trusov, A.~I.~Veselov }\\      
{\small \it ITEP, Moscow, Russia}
\end{center}

\vspace*{6mm}

\begin{abstract}
QCD string model formulated in the framework of the Field
Correlator Method (FCM) in QCD is employed to calculate the masses
of $\Sigma_c$, $\Xi_c$ and recently observed at Tevatron
$\Sigma_b$, $\Xi_b$ baryons and their orbital excitations.

The auxiliary field formalism allows one to write a simple local
form of the effective Hamiltonian for the three quark system,
which comprises both confinement and relativistic effects, and
contains only universal parameters: the string tension $\sigma$,
the strong coupling constant  $\alpha_s$, and the bare (current)
quark masses $m_i$. We calculate the hyperfine splitting with
account of the both perturbative and non--perturbative spin-spin
forces between quarks in a baryon. For the orbital excitations we
estimate the string correction for the confinement potential~ -
~the leading correction to the contribution of the proper inertia
of the rotating string. This correction lowers the masses of the
P-states by ~ 50 MeV. We find our numerical results to be in good
agreement with experimental data.

\end{abstract}

\section{Introduction}
The spectroscopy of $c$ and $b$ baryons has undergone a great
renaissance in recent years. New results have been appearing in
abundance as a result of improved experimental techniques
including information on states made of both light {($u,\,d,\,s$)}
and heavy $(c,\,b)$ quarks. Before 2007, the only one baryon with
a $b$ quark, the isospin-zero $\Lambda_b^0$, was known.  Now,
we have the isospin one $\Sigma_b$, $\Sigma_b^{*}$ baryons and
$\Xi_b$. The CDF Collaboration has seen the states
$\Sigma_b^{\pm}$ and $\Sigma_b^{*\pm}$ \cite{cdf_1},
while D\O~ \cite{do} and CDF  \cite{cdf_2} have observed the
$\Xi_b^-$ (for the review see \cite{M07}).

On theoretical side there are many results on heavy baryon masses
from different approaches, see {\it e.g.} \cite{RP2007} and
references therein.
In the present paper we use the field correlator method (FCM)
\cite{DS} to calculate the masses of the $S$ wave baryons
containing $c$ and $b$ quarks and orbitally excited $P$ wave
states that will be experimentally accessible in the future.
\section{Effective Hamiltonian and baryon masses in MFC}

The key ingredient of the FCM is the use of the auxiliary fields
(AF) initially introduced in order to get rid of the square roots
appearing in the relativistic Hamiltonian, see \cite{NSV} and
references therein. Using the AF formalism allows to write a
simple local form of the Effective Hamiltonian (EH) for the three
quark system \cite{Si2003}
\begin{equation}
\label{eq:H} H=\sum\limits_{i=1}^3\left(\frac {m_{i}^2}{2\mu_i}+
\frac{\mu_i}{2}\right)+H_0+V,
\end{equation}
where $H_0$ is the kinetic energy operator, $V$ is the sum of the
string potential and a one gluon exchange potential, $m_i$ are the
bare quark masses,
and $\mu_i$ are the {\it constant} AF which are eventually treated
as variational parameters. Such an approach allows one a very
transparent interpretation of AF: starting from bare quark masses
$m_i$ one arrives at the dynamical masses $\mu_i$ which appear due
to the interaction and have the meaning of quark  constituent
masses. The string potential is \be\label{eq:string} V_{Y}({\bf
r}_1,\,{\bf r}_2,\,{\bf r}_3)\,=\,\sigma\,r_{min},\ee  where
$\sigma$ is the string tension and  $r_{min}$ is the minimal
length corresponding to the Y--shaped string configuration. The
one--gluon exchange potential is \be\label{eq:coulomb}
V_{C}\,=\,-\frac{2}{3}\,\alpha_s\,\sum_{i<j}\,\frac{1}{r_{ij}},\ee
$r_{ij}$ being the interquark distances. In Eq. (\ref{eq:coulomb})
$\alpha_s\,=\,0.39$ is the freezing value of the strong coupling
constant.

The mass $M_B$ of a $S$--wave baryon is given by \be\label{eq:M_B}
M_B\,=\,M_0\,+\,\Delta E_{HF},\ee where $\Delta E_{HF}$ is the
perturbative hyperfine correction, and
\begin{equation}
\label{eq:M_B0}M_0\,=\,\sum\limits_{i=1}^3\left(\frac
{m_{i}^2}{2\mu_i\,}+
\,\frac{\mu_i}{2}\right)\,+\,E_0(\mu_i)\,+\,C,\end{equation}
$E_0(\mu_i)$ being the energy eigenvalue of the Shr\"{o}dinger
operator $H_0\,+\,V$.
The constant $C$ in (\ref{eq:M_B0}) is the calculable quark
self-energy correction \cite{S2001} which is created by the color
magnetic moment of a quark propagating through the vacuum
background field:
\begin{equation} \label{self_energy}
C\,=\,-\frac{2\sigma}{\pi}\,\sum\limits_i\frac{\eta(t_i)}{\mu_i},
\,\,\,\,\,t_i\,=\,m_i/T_g,\end{equation}  where $1/T_g$ is the
gluonic correlation length. We use $T_g\,=\,1$ GeV. The numerical
factor $\eta(t)$ arises from the evaluation of the integral
\be \eta(t)= t\int^\infty_0 z^2\, K_1(tz)\,
e^{-z}\,dz,\label{eq:eta} \ee where $K_1$ is the McDonald
function.  This correction adds an overall negative constant to
the hadron masses.

The hyperfine splitting $\Delta\,E_{HF}$ is given by
\be\Delta\,E_{HF}\,=\,<\Psi|H_{HF}\,|\Psi>
 \,=\,\sum\limits _{i<j}\,\frac{\bsl\sigma_i\,
             \bsl\sigma_j}{\mu_i\mu_j}\,\left(\frac{4\pi\alpha_s}{9}\,\left\langle\,
             \delta({\bsl r}_{ij})\right\rangle\,+\,\frac{\sigma\,T_g^2}{4\pi}\,\langle r_{ij}\cdot
             K_1(T_g\,
              r_{ij})\rangle\right),\label{eq:HF}
\ee and $\Psi$ is the eigenfunction of the Hamiltonian
$H_0\,+\,V$. The first term in (\ref{eq:HF}) is the standard
color-magnetic interaction in QCD \cite{deRuhula}, while the
second term, proportional to the string tension $\sigma$, was
derived in Ref. \cite{S2002}.
\section{The string corrections}

For $P$--wave baryons we
explicitly calculate the so--called string correction \cite{DKS}.
Recall that the string potential $V_Y({\bf r}_1,\,{\bf r}_2,\,{\bf
r}_3)$ in Eq. (\ref{eq:string}) represents only the leading term
in the expansion of the QCD string Hamiltonian in powers of
angular velocities .
The leading correction in this expansion is known as a string
correction. This is the correction totally missing in relativistic
equations with local potentials. Its sign is negative, so the
contribution of the string correction lowers the energy of the
system, thus giving a negative contribution to the masses of
orbitally excited states, leaving the S--wave states intact.

So far the string correction was calculated for the orbitally
excited mesons \cite{KNS} and hybrid charmonium states
\cite{KN2008}. For a baryon with the genuine string junction point
the calculation of the string correction is a very cumbersome
problem. The calculations are greatly simplified, however,  if the
string junction point is chosen as coinciding with the
center--of--mass coordinate ${\bsl R}_{\text cm}$. In this case
the complicated string junction potential is approximated by a sum
of the one--body confining potentials. The accuracy of this
 approximation for the $P$--wave baryon states
 is better than 1$\%$ \cite{NSV}.
 Letting ${\bsl R}_{cm}\,=\,0$ we arrive
to the string potential in the form
 \be\label{eq:V_string} V_{\rm string}^{\rm CM}\,=\,\sigma \,\sum_i\,|\bsl{r}_i|\int^1_0 d\beta_i\sqrt{1-{\bsl l}_i^2}
, \ee where \be{\bsl l}_i\,=\,\frac{\beta}{|{\bsl r}_i|}\,[{\bsl
r}_i\,\times\,\dot{\bsl r}_i]\,=\,\frac{\beta}{m_i\,|{\bsl
r}_i|}\,[{\bsl r}_i\,\times\,{\bsl
p}_i]\,\Rightarrow\,-i\,\frac{\beta}{m_i\,|{\bsl r}_i|}\, [{\bsl
r}_i\,\times\,{\bsl\nabla}_i].\ee Expanding the square roots in
Eq. (\ref{eq:V_string}) in powers of the angular velocities ${\bsl
l}_i^2$ and keeping only the first two terms in this expansion one
obtains
 \be V_{\rm
string}^{\rm CM}\,\approx\,\,\sigma\,\sum_i\,|{\bsl
r}_i|\,\int\limits_0^1\,d\beta\,(1\,-\,\frac{1}{2}\,{\bsl
l}_i^2)\,=\,\sigma\,\sum_i\,\left(\,|{\bsl
r}_i|\,+\,\frac{1}{6}\cdot\frac{1}{m_i^2\,|{\bsl r}_i|}({\bsl
r}_i\,\times\,{\bsl\nabla}_i)^2\right) \ee and
 \be\label{eq:Delta_M}\Delta M_{\rm
string}\,=\,-\,\frac{\sigma}{6}\,<\Psi|\sum_i\,\frac{({\bsl
r}_i\times{\bsl p}_i)^2}{m_i^2\,r_i}\,|\Psi>,\ee where $\Psi$ is
an eigenfunction of the Hamiltonian (\ref{eq:H}).  Further details
and numerical results will be  given in Sec. \ref{sect:results}.

\section{Results and discussion}\label{sect:results}
The dynamics of the $ud$ pair plays a relevant role, being mainly
responsible for the spin splitting in the strange sector. A
similar contribution is expected for charmed and bottom baryons.
Estimates of the one-pion exchange contribution to the baryon mass
give $\sim\,$-200 MeV both for $\Lambda$ and $\Lambda_b$ \cite{V}.
Because our approach misses the chiral physics effects we
calculate in this work the masses of the $\Sigma$ and $\Xi$ states
which are affected by the chiral dynamics only slightly.
 Following our previous analysis \cite{DNV} we
employ the hyperspherical method to calculate the energy
eigenvalues $E_0(\mu_i)$, the constituent quark masses $\mu_i$ and
the zero-order baryon masses $M_{0}$ in Eq. (\ref{eq:M_B0}).
For $nnQ$ baryons (with $Q$ standing for either $c$ or $b$) we use
the basis in which the heavy quark $Q$ is singled out as quark $3$
but in which the non-strange quarks are still antisymmetrized.
The $nnQ$ basis states diagonalize the confinement problem with
eigenfunctions that correspond to separate excitations of the
light and heavy quarks (${\rho}$\,- and ${\lambda}$\, excitations,
respectively). In particular, excitation of the $\bsl{\lambda}$
variable unlike excitation in $\bsl{\rho}$ involves the excitation
of the ``odd'' $Q$ quark . In the $\Xi_Q$ the $n$ and $s$ quarks
are approximately in a state with $S\,=\,0$, while another heavier
state $\Xi'_Q$ is expected in which the $n$ and $s$ quarks mainly
have $S\,=\,1$. Both have total $J\,=\,1/2$. The effect of
$\Xi\,-\,\Xi'$ mixing due to the spin-spin interaction is
negligible \cite{KKLR}. There is also a state $\Xi_Q^*$ expected
with total $J\,=\,3/2$.

We use the approximation $K\,=\,K_{\rm min}$, where $K_{\rm
min}\,=\,0$ for $L\,=\,0$ and $K_{\rm min}\,=\,1$ for $L\,=\,1$,
where $K$ is the grand orbital momentum. The accuracy of this
approximation was checked in Ref. \cite{NSV}. Then the reduced
wave function $u_{\nu}(x)$ ( $\nu\,=\,0$~ for $L\,=\,0$,~and
$\nu\,=\,\rho,\,\lambda$ for $L\,=\,1$), satisfies the equation
\be\label{eq:se}\frac{d^2
u_{\nu}(x)}{dx^2}\,+\,2\left(E_0\,-\,\frac{(K+\frac{3}{2})(K+\frac{5}{2})}{2\,x^2}\,-
\,V_{\rm Y}^{\nu}(x)\,-\,V_{\rm
C}^{\nu}(x)\right)u_{\nu}(x)\,=\,0,\ee where \be
x^2\,=\,\sum_i\,\mu_i\,({\bsl r}_i\,-\,{\bsl
R}_{cm})^2\,=\,\frac{\mu_1\,\mu_2}{M}\,r_{12}^2\,+\,\frac{\mu_2\,\mu_3}{M}\,r_{23}^2\,+\,
\frac{\mu_3\,\mu_1}{M}\,r_{31}^2,\ee \[\qquad M=\mu_1+\mu_2+\mu_3,
\],
\be\label{eq:b} V_{\rm Y}^{\,\nu}(x)\,=\,\int
\,|Y_{\nu}\,(\theta,\chi)|^2\,V_{\rm Y}({\bf r}_1,\,{\bf
r}_2,\,{\bf r}_3)\,d\Omega\,=\,
\sigma\, b_{\nu}\,x,\ee and\be\label{Coulomb}V_{\rm
Coulomb}^{\,\gamma}(x)\,=\,-\,\frac{2}{3}\,\alpha_s\,\int
\,|Y_{\nu}\,(\theta,\chi)|^2\,\sum_{i\,<\,j}\,\frac{1}{r_{ij}}\,\,\,d\Omega\,=\,
-\,\frac{2}{3}\,\alpha_s\,\frac{a_{\nu}}{x},\ee The hyperspherical
harmonics are \bea
Y_0\,=\,\sqrt{\frac{1}{\pi^3}}\,\,\,\,\,(K\,=\,0),\,\,\,\,\,\,\,\,\,\,\,\,\,\,\,\,\,\,\,
\bsl{Y}_{\rho}\,=\,\sqrt{\frac{6}{\pi^3}}\,\frac{\bsl{\rho}}{R}\,,\,\,\,\,\,\,\,
\bsl{Y}_{\lambda}\,=\,\sqrt{\frac{6}{\pi^3}}\,\frac{\bsl{\lambda}}{R}\,\,\,\,\,(K\,=\,1),\eea
where $\bsl\rho$, $\bsl\lambda$ are the Jacobi coordinates, and
$R^2\,=\,{\bsl\rho}^2\,+\,{\bsl\lambda}^2.$

Explicit expressions for the constants $a_{\nu}$ in
(\ref{Coulomb}) and the two--dimensional integrals defining the
constants $b_{\nu}$ in Eq. (\ref{eq:b})
 are written in Ref. \cite{DNV}).

For the S-wave baryons we calculate the hyperfine splittings from
Eq. (\ref{eq:HF}).
We employ $m_n\,=\,7\,$ MeV (with $n$ standing for either $u$ or
$d$) and the strange quark mass $m_s\,=\,185$ MeV found previously
from the fit to $D_s$ spectra. However, our predictions need an
additional input for the bare quark masses $m_c$ and $m_b$. These
were fixed from the masses of $\Sigma_c$ and $\Sigma_b$,
respectively, $m_c\,=\,$ 1359 MeV and $m_b\,=\,$ 4712 MeV.
\begin{table}[t]
 \caption{Heavy Baryons with $L\,=\,0$.
The underlined masses have been used to fix $m_c$ and $m_b$. The
experimental baryon masses are for the isospin averaged states.
All masses are in units of MeV.} \vspace{5mm}

\centering
\begin{tabular}{ccccccccc} \hline\hline\\
Baryon&~~$\mu_n$&
~~$\mu_s$&~~$\mu_h$&$M_0$&~$\Delta\,E_{HF}$~&~$M$~&$M_{\rm exp}$
\\  \\\hline\hline\\
$\Sigma_c$~~~&~~470~~&&~~1455~~&~~2479~~&-25&\underline {2454}&2455\\
$\Sigma_c^*$~~&~~470~~&&~~1455~~&~~2479~~&43&2522&2520\\
$\Xi_c$~~~&~~476&~~522&~~1458 &2519&-59&2460&2471\\ \\ \hline\\
$\Sigma_b$~~~&~~509&&~~4749&5806&2&\underline{5808}&5810\\
$\Sigma_b^*$~~&~~509&&~~4749&5806&+27&5833&5830\\
$\Xi_b$~~~&~~514&~~615&~~4751&5844&-53&5791&5790\\
\\ \hline\hline
\end{tabular}
 \vspace{1mm}

\label{L=0}
\end{table}


The result of the calculation of the $S$ wave states is given in
Table \ref{L=0}. In this Table we also present the dynamical quark
masses $\mu_n$, $\mu_s$ and $\mu_Q$ (columns 2--4) for various
baryons ($Q$ standing for either $c$ or $b$). The latter are
computed solely in terms of the bare quark masses, $\sigma$ and
$\alpha_s$ and marginally depend on a baryon. We also display the
zero--order masses ( calculated without the HF corrections from
Eq. (\ref{eq:M_B0}) (column 5), the HF correction (column 6) and
the total baryon masses (\ref{eq:M_B}) (column 7). Experimental
masses (column 8) are from Ref. \cite{PDG08}.

The result show good agreement between data and theoretical
predictions. In particular, the hyperfine splitting between
$\Xi_c^*$ and $\Xi_c'$
 is found to be 69 MeV that
agrees with the experimental value ($\sim$ 70 MeV), while the
predicted mass difference $\Xi_b^*\,-\,\Xi_b'\,=\,$ 26 MeV agrees
with the finding of Ref. \cite{KKLR}. However, our perturbative
calculations do not reproduce the observed $\Xi_c'$ - $\Xi_c$ mass
difference. The large hyperfine splitting between axial and scalar
$ns$ diquarks is usually described by the smeared
$\delta$-function that requires additional model-dependent
assumptions about the structure of interquark forces.
\begin{table}[t]
 \caption{Heavy Baryons. $L\,=\,1$.
The bare quark masses are the same as in Table \ref{L=0}.}
\label{tab:L=1} \vspace{2mm}

\centering
\begin{tabular}{ccccccccc} \hline\hline\\
Baryon&~~${\bf L}_{\alpha}$ &~~$\mu_n$&~~ $\mu_s$&~~$\mu_h$&~~
$E_{0}$&$M_0$&$\Delta M_{\rm string}$& $M_B$\\ \\ \hline\hline\\
$nnc$&${\bf 1}_{\rho}$&536&&1452&1397&2920&-48&2872\\  $nnc$&${\bf
1}_{\lambda}$&495&&1491&1377&2832&-36&2796
\\
$nsc$& ${\bf 1}_{\rho}$&542&582&1455&1372&2954&-45&2909\\
$nsc$& ${\bf 1}_{\lambda}$&497&544&1494&1353&2867&-33&2834\\
\\

\hline\\
$nnb$&${\bf 1}_{\rho}$&570&&4746&1294&6240&-46&6194\\  $nnb$&${\bf
1}_{\lambda}$&540&&4764&1234&6132&-41&6091
\\
$nsb$&${\bf 1}_{\rho}$&574&615&4748&1271&6272&-43&6229
\\
$nsb$&${\bf 1}_{\lambda}$&542&588&4765&1211&6164&-38&6126
\\ \\
\hline\hline

\end{tabular}
\vspace{2mm}

\end{table}

A similar calculations were performed for the P-wave
orbitally-excited states, see Table \ref{tab:L=1}.  Our basis
states diagonalize the confinement problem with eigenfunctions
that correspond to separate excitations of the light and heavy
quarks ($\rho$ - and $\lambda$ - excitations, respectively).
Excitation of the $\lambda$ variable unlike excitation in $\rho$
involves the excitation of the ``odd'' heavy quark.

Explicit calculation \cite{DN} of the string correction
(\ref{eq:Delta_M}) yields: \be\Delta
M_{\rho}\,=\,-\,\frac{64\,\sigma}{45\pi}\,\frac{1}{\sqrt{M}\,(\mu_1\,+\,\mu_2)}\left(
\frac{\sqrt{\mu_2\,+\,\mu_3}}{\mu_1^{3/2}}\mu_2\,+\,\frac{\sqrt{\mu_1\,+\,\mu_3}}{\mu_2^{3/2}}\mu_1
\right)\,\gamma_{\rho},\ee
 \be\Delta
 M_{\lambda}=-\frac{64\sigma}{45\pi}\,\frac{\mu_3}{M^{3/2}(\mu_1+\mu_2)}
 \left(
\left[\frac{\mu_1\,+\,\mu_2}{\mu_3}\right]^{5/2}\,+\,\sqrt{\frac{\mu_2\,+\,\mu_3}{\mu_1}}\,+\,
\sqrt{\frac{\mu_1\,+\,\mu_3}{\mu_2}}\,
\right)\,\gamma_{\lambda},\ee
where \be\gamma_{\nu}\,=\,\int\limits_0^{\infty}
 \frac{{u}_{\nu}^2(x)}{x}dx\ee
 For states with one unit of orbital angular momentum
between $Q$ quark and the two light quarks we obtain
$M(\Sigma_{c}) = 2796$ MeV, $M(\Xi_{c}) = 2834$ MeV,
$M(\Sigma_{b}) = 6091$ MeV, and $M(\Xi_{b}) = 6126$ MeV, while the
states with one units of orbital momentum between the two light
quarks are typically $\sim$ 100 MeV heavier. Note that zero order
results of Table \ref{tab:L=1} do not include the spin
corrections.
A more complete
analysis will be given elsewhere.

\section{Conclusions}
We have calculated the masses of heavy baryons systematically
using the string model in QCD and the  color-magnetic interaction.
There are three main points in which we differ from other
approaches to the same problem based on various relativistic
Hamiltonians and equations with local potentials. The first point
is that we do not introduce the constituent mass by hand. On the
contrary, starting from the bare quark mass we arrive to the
dynamical quark mass that appears due to the interaction. The
second point is that for the first time we calculate the hyperfine
splitting with account of the nonperturbative spin-spin forces
between quarks in a baryon. Finally, for the first time we
calculate the string correction for the $P$--wave states of heavy
baryons.
\vspace{1cm}

This work was supported by the RFBR grants  06-02-17120,
08-02-00657, 08-02-00677, and by the grant for scientific schools
NSh.4961.2008.2 .


\end{document}